\documentclass[sigconf]{acmart}

\AtBeginDocument{%
  }
\usepackage{natbib}
\usepackage{graphicx}
\usepackage{subfigure}
\usepackage{textcomp}
\usepackage{xcolor}
\usepackage{algpseudocode}
\usepackage{algorithm}
\usepackage{amsthm}
\usepackage{subfigure}
\usepackage{wrapfig}
\usepackage{mathtools}
\usepackage{multirow}
\theoremstyle{plain}
\newtheorem{theorem}{Theorem}

\theoremstyle{definition}
\newtheorem{definition}[theorem]{Definition}

\theoremstyle{remark}

\usepackage{enumitem}
\usepackage{xcolor}
\usepackage{pifont}
\usepackage{xspace}
\usepackage{cleveref}
\newcommand{\ourmethod}{\textsc{ProtoHAIL}\xspace}
\newcommand{\eg}{\emph{e.g.}\xspace}
\newcommand{\ie}{\emph{i.e.}\xspace}
\newcommand{\etaldot}{\emph{et al.}\xspace}
\newcommand{\code}[1]{\texttt{#1}}

\newcommand\Mycomb[2][^n]{\prescript{#1\mkern-0.5mu}{}C_{#2}}

\acmBooktitle{KDD '23}

\begin{document}
\pagestyle{plain}
\title{Risk-aware Adaptive Virtual CPU Oversubscription in Microsoft Cloud via 
Prototypical Human-in-the-loop
Imitation Learning}

\author{Lu Wang$^*$, Mayukh Das$^*$, Fangkai Yang$^*$, Junjie Sheng$^*$, Bo Qiao$^*$, Hang Dong$^*$, Si Qin$^*$,\\ Victor Rühle$^*$, Chetan Bansal$^*$, Eli Cortez$^+$, \'I\~nigo Goiri$^+$, Saravan Rajmohan$^*$, Qingwei Lin$^*$, Dongmei Zhang$^*$}
\affiliation{%
 \vspace{0.5em}\institution{
     Microsoft$^*$ 
     \hspace{0.3em} 
     Microsoft Azure$^+$ 
     \hspace{0.3em}
 }
 \country{}
}

\begin{abstract}
Oversubscription is a prevalent practice in cloud services where the system offers more virtual resources, such as virtual cores in virtual machines, to users or applications than its available physical capacity for reducing revenue loss due to unused/redundant capacity. While oversubscription can potentially lead to significant enhancement in efficient resource utilization, the caveat is that it comes with the risks of overloading and introducing jitter at the level of physical nodes if all the co-located virtual machines have high utilization. Thus suitable oversubscription policies which maximize utilization while mitigating risks are paramount for cost-effective seamless cloud experiences. Most cloud platforms presently rely on static heuristics-driven decisions about oversubscription activation and limits, which either leads to overloading or stranded resources. Designing an intelligent oversubscription policy that can adapt to resource utilization patterns and jointly optimizes benefits and risks is, largely, an unsolved problem. We address this challenge with our proposed novel HuMan-in-the-loop Protoypical Imitation Learning (ProtoHAIL) framework that exploits approximate symmetries in utilization patterns to learn suitable policies. Also, our human-in-the-loop (knowledge-infused) training allows for learning safer policies that are robust to noise and sparsity. 
Our empirical investigations on real data show orders of magnitude reduction in risk and significant increase in benefits (saving stranded cores) in Microsoft cloud platform for 1st party (internal services). 
\end{abstract}




\maketitle

\begin{figure}[t]
  \centering
    \includegraphics[width=0.4\textwidth]{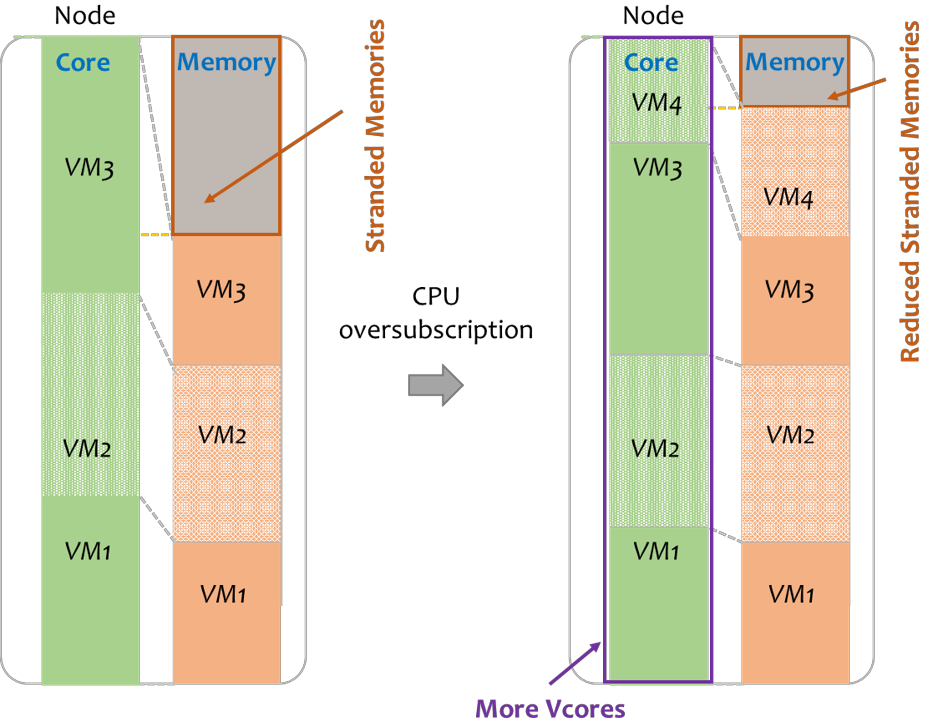}
  \caption{VCPU oversubscription.}
  \label{fig:vcpuoversub}
  \vspace{-6mm}
\end{figure}
\section{Introduction}

The term \textit{Oversubscription} is widely used to characterize scenarios where
a system offers more resources or services to users or entities than its available capacity, assuming not all users would simultaneously or fully utilize the allocated capacity. 

In cloud services, requested resources are terms of Virtual Machines (VMs) each having some quantum of virtual resources such as virtual cores/memory etc. and these VMs are, in turn, hosted on Physical Machines (PMs). Oversubscription here assigns fewer resources for a VM than what was requested, assuming that the VM will be allowed to use extra resources beyond its assigned limit if needed (as shown in Figure 1). This scheme allows the platform to leverage unused physical capacity by packing more VMs as an effective way to avoid unnecessary resource waste and maximize profits for the internal teams~\cite{baset2012towards, breitgand2012improving}.
CPU bottlenecks are more severe and prevalent than memory and network in cloud~\cite{mahapatra1999processor}. As illustrated in the representative example in  Figure~\ref{fig:vcpuoversub}, three VMs are placed in the same node, and the CPU dimension is fully packed with unused memory, \ie, stranded memory (grey box). When oversubscribing, additional VMs can be allocated on this node reducing stranded memory. Note that, virtual CPUs (vCPU) represent the share of the underlying physical CPU assigned to VMs and are the sellable billing units in cloud platforms. So, we focus on vCPU oversubscription. 
However, designing a suitable oversubscription policy is both critical and challenging\textemdash~once a system is oversubscribed, overloading or under-utilization may happen~\cite{baset2012towards} at any point.

Forecasting the users' demand and utilization behaviors at the correct granularity and cadence to design adaptive policies is difficult.
Aggressive oversubscription unfairly penalizes an uncertain amount of users who cannot access the resources, also known as \textit{overloading} in cloud~\cite{williams2011overdriver},
leaving them unsatisfied risking 
business reputation and revenue losses through monetary compensations~\cite{wittman2014low}. Again, conservative policies result in unused resources and capacity, leading to inefficient resource usage or even wastage. 
We address the problem of adaptable vCPU oversubscription for Microsoft's internal Cloud system by minimizing overloading risks while maximizing utilization. Most existing research in this space does not offer any principled or generalized formulation and restricts the problem to their specific scenarios. For instance, oversubscription in cloud is formulated as a variant of the online bin-packing problem with constraints~\cite{baset2012towards,breitgand2012improving, householder2014cloud}, addressing the resource allocation policy instead of designing an oversubscription policy. Others focus on usage migration to mitigate overload situations~\cite{wang2018energy, li2019adaptive}.
Yet, a generalized oversubscription policy, with essentially competing objectives of efficiently utilizing unused capacity while minimizing overload risks, is still underexplored. 

In this paper, we pose the risk-aware adaptive oversubscription question as a sequential decision-making problem with constraints or resource limits. 
Predicting future utilization behaviors given historical observations through traditional supervised learning approaches is insufficient since they are unaware of the interactions between the users and the environments~\cite{de2019causal}. 
Again, in traditional reinforcement learning with constraints~\cite{garcia2015comprehensive}, it is challenging to balance different, possibly competing, objectives 
with convergence guarantees due to non-convexity of the  ~\cite{achiam2017constrained, paternain2019constrained,mazyavkina2021reinforcement}.
Imitation learning (IL), on the other hand, can be leveraged to solve MDP constraint problems~\cite{hussein2017imitation} where the expert's policy fulfills the constraints by nature. In particular, we propose a prototypical imitation learning method (\ourmethod) that learns to take actions by a set of learned \textit{prototypes}, where prototype is a data instance that is representative of an equivalence class of expert trajectories~\cite{kim2016examples,molnar2020interpretable}.Consequently our approach can leverage approximate symmetries in utilization patterns, allowing us to learn adaptive policies for any granularity. One caveat is that the utilization data is usually noisy, incomplete, or sparse, resulting in sub-optimal prototypes and policies. Thus, facilitated by the interpretability of prototypes, we leverage human-in-the-loop training to learn safer and better policies. 

We make the following major contributions: (1) We propose a novel prototypical imitation learning approach to solve the vCPU oversubscription rate prediction problem for Microsoft's internal cloud to maximize utilization efficiency and minimizing risk; (2) To mitigate challenges of noisy, sparse trajectories, we propose an efficient active human-in-the-loop (HITL) training technique; (3) We show, via extensive evaluations on Microsoft 365's internal Cloud platform, how \ourmethod learns effective risk-aware and adaptable oversubscription policies; (4) We also show how our \ourmethod seamlessly generalizes to any oversubscription problem in other domains beyond cloud services, via additional experiments on a novel airline ticket overbooking dataset {curated from static periodic reports of  U.S. Department of Transportation (DOT)}. The code of our method can be found here\footnote{https://github.com/Joywanglulu/Oversubscription.git}.

\section{Preliminary}
\label{Preliminary}
\subsection{Prototype theory}
Prototype theory emerged in 1971 with the work of psychologist Eleanor Rosch~\cite{rosch1973natural} and is known as the ``Copernican revolution" in the theory of categorization. In prototype theory, any given concept in any given language has a real-world example that best represents this concept. For instance, as an example of the concept \textit{fruits}, an apple is more frequently cited than a durian. This theory claims that the presumed natural prototypes are central tendencies of the categories. Prototype theory has also been applied in 
machine learning~\cite{kim2016examples}, where a prototype is defined as a data instance that is representative of all the data~\cite{molnar2020interpretable}. There are many approaches to find prototypes in the data. Any clustering algorithm that returns actual data points as cluster centers would qualify for selecting prototypes. For example, Harel \etaldot~\cite{harel2018accelerating} utilizes a generative model to cluster the data and consider the central points as prototypes. ProSeNet~\cite{ming2019interpretable} introduces a set of constraints to learn simplicity, diversity, and sparsity prototypes with a sequence encoder. However, to our best knowledge, there is no prior work leveraging prototype theory in IL. In fact, leverage prototypes into imitation learning, can help us understand the different types of expert policies. 

\textbf{Motivation:} In cloud system, we can consider the users' CPU usage rate as expert policy of oversubscription. So different equivalence classes of CPU usage patterns can be considered as different prototypes. Once the prototypes are learnt, oversubscription action is predicted via referring the nearest prototype as per input and context. There are two main advantages of imitation learning with prototypes: (i) it helps learn interpretable policies which allows for naturally querying the human's feedback to reduce risk. (ii) It can results in more effective oversubscription policy with learning from the prototypes. Thus, in this paper, we learn prototypical options to understand the experts' policy via case-based interpretability. 

\subsection{Human-in-the-loop learning}   
\textbf{Motivation:} In oversubscription, systematic noise in utilization or behavior trajectories are prevalent, resulting in sub-optimal model learning. For instance, the vCPU utilization data has sample sparsity and variance\textemdash~resulting in incorrect policy or pattern clusters.
Knowledge-guided (Human-in-the-loop) systems allow us to leverage knowledge from domain experts beyond what can be gleaned from data to learn better models. Typically, there are two modes in HITL\textemdash~ (1) passive priors or constraints, (2) active feedback from domain experts during the training on model/outputs. We adopt the active mode as it is difficult to design appropriate priors/constraints in prototypical IL without visibility into what or how prototypes are learned. Also, active knowledge has richer information than upfront priors \cite{neider2021advice, pmlr-v87-brown18a, DENG2020101656}. 
While knowledge representation is also an important aspect, such as explanations~\cite{minton1989explanation}, preference rules~\cite{odom2015active,das2021human}, implicit pseudo-labeling~\cite{jiang2018mentornet,goldberger2016training,patrini2017making,miyato2018virtual,lee2013pseudo}, it is not our focus/scope. We represent feedback with simple indicators, such as upvote, downvote, merge, split, etc. Knowledge-guided learning is equivalent adding an inductive bias to nudge sub-optimal models (distributions) learned from noisy observations towards, possibly unknown, optimal distribution; $\pi^*(\cdot) = \pi_{data}(\cdot)\oplus \pi_{know} (\cdot)$, where $\pi^*$ is the optimal distribution. Some approaches formulate this directly as a mixture distribution $\pi^*=\alpha \pi_{data}+(1-\alpha) \pi_{know}$, with trade-off $\alpha$, while others use implicit regularization or constraints on the loss function
\cite{fung2002knowledge,kunapuli2013guiding}. 
We adopt a different formulation from KCLN \cite{das2021human}.

%

\section{Method}
\label{method}
\subsection{Problem Formulation}
We formulate the oversubscription problem in Cloud as a prototypical imitation learning problem. In cloud system, we can consider the CPU usage rates as expert trajectories of oversubscription. Then different classes of CPU usage patterns can be considered as different prototypes. The goal of prototypical imitation learning is to identify these users' prototypes, then predict ideal oversubscription via referring the nearest prototypes based on the current context. 

\begin{definition}[Prototypical Imitation Learning]
We formally define prototypical imitation learning as a kind of imitation learning that learns an oversubscription policy by aligning with a reference prototype (prototypical trajectories) from the experts' policy. Specifically, each prototype intuitively represents equivalence classes of patterns for decision-making references and analogical explanations. Prototypical imitation learning learns a metric space in which decision-making can be performed by computing the distance to the prototype policies of the expert. Then the prediction of new action can be derived and explained by the closest prototype trajectories. 

\end{definition}


\subsection{Overview of Prototypical Imitation Learning}

\begin{figure}
    \centering
    \includegraphics[width=0.48\textwidth]{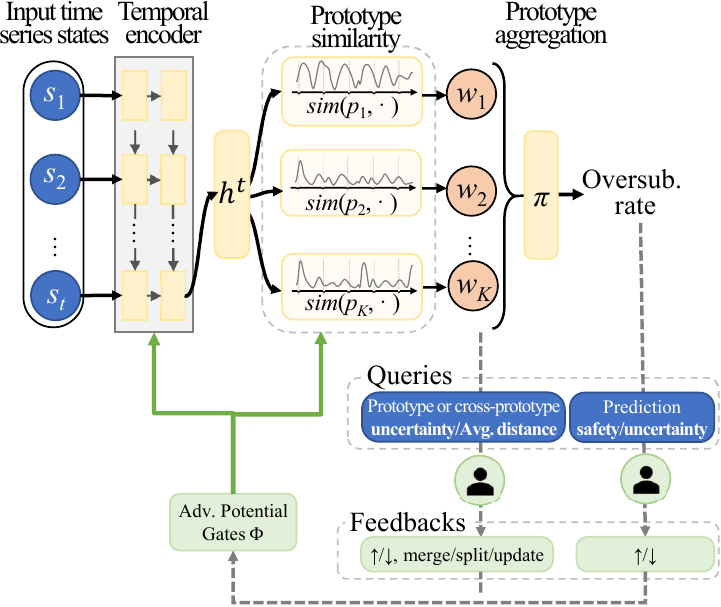}
  \caption{Model overview}
  \label{fig:model}
  \vspace{-10pt}
\end{figure}

The basic architecture of \textbf{\underline{Proto}}typical \textbf{\underline{H}}um\textbf{\underline{A}}n-in-the-loop \textbf{\underline{I}}mitation \textbf{\underline{L}}earning (\ourmethod) is shown in Figure~\ref{fig:model}. 
There are three main components: (1) discovering prototypes, where we classify the trajectories into $K$ groups and learn a prototypical representation $p_k$; (2) learning an action policy by aligning with its similar prototypes at different states; (3) obtaining feedback from the human-in-the-loop to reduce the risk of the policy.

For a given input trajectory till timestamp $t$, $\tau_t=\{(s_1,a_1,...,\\s_{t-1}, a_{t-1},s_t)\}$, the trajectory encoder $f$ maps $\tau_t$ into an embedding vector $h=f(\tau_t)$, $h\in \mathbb{R}^m$. The encoder could be any sequence encoder model, \eg, LSTMs or Transformers. 
The prototype layer $p$ contains $K$ prototype embeddings $\mathcal{P}=\{p_1,\dots,p_K\}$, where each $p_k\in\mathbb{R}^m$ have the same length as $h$. This layer scores the similarity between $h$ and each prototype $p_k$. We consider the $L_2$ distance metric as the similarity function for simplicity. With the computed similarity vector $[sim(f(\tau_t),p_1),\dots, sim(f(\tau_t),p_K)]$, the policy layer $\pi$ computes the action with a linear layer with sigmoid activation.
In the end, an HITL module is leveraged to refine \ourmethod for better interpretability and performance by validating and refining the prototypes.

As for the oversubscription problem, $s_t$ indicates the features of resources including the resource requirements and the resource allocations. The action $a_t\in[0,1]$ indicates the oversubscription rate defined as the ratio of allocated resources over requested resources.



\subsection{Prototype Discovery}

The \textit{prototype} is defined as the representative instance $\tau^{E,k}$ selected from a class of expert trajectories. 
Specifically, the trajectory encoder
is shared for encoding $\tau_t$ and $\tau^{E,k}$ to the same embedding space, \ie, $p_k=f(\tau^{E,k})$ and $h=f(\tau_t)$. Note that $\tau^{E,k}$ is the full expert trajectory and we omit the time subscript.
We consider three characteristics as the objective for learning prototypes, \ie, representative, diversity, and interpretability. 


For \textbf{representative}, we aim to learn prototypes that can well represent a subset of trajectories with the regularization term $\mathcal{L}_{rep}$. It encourages a clustering structure in the embedding space by minimizing the $L_2$ distance between an encoded trajectory and its nearest prototype embedding: 
\begin{align}
    \mathcal{L}_{rep} = \frac{1}{K}\sum^K_{k=1}\frac{1}{|D_k|} \sum_{\tau\in D_k} || p_k - f(\tau) ||^2_2,
\end{align}
where $D_k$ indicates a subset of trajectories that could be represented by $p_k$.

Now, to improve the \textbf{diversity} and reduce redundancy among prototypes, the term  $\mathcal{L}_{div}$  penalizes the prototypes that are close to each other:
\begin{align}
    \mathcal{L}_{div} = -\frac{1}{\Mycomb[K]{2}}\sum^K_{i=1}\sum^K_{j=i+1} || p_i - p_j ||^2_2,
\end{align}
where $\Mycomb[K]{2} = \frac{K!}{2!(K-2)!}$.

To give \textbf{interpretability}, we assign one expert trajectory instance $\tau^{E,k}$ as one prototype via $\mathcal{L}_{int}$. Then, each prototype embedding can be explained by a real-world instance.
\begin{align}
    \mathcal{L}_{int} = \frac{1}{K}\sum^K_{k=1} || p_k - f(\tau^{E,k}) ||^2_2, 
\end{align}

where $\tau^{E,k}$ is the nearest expert trajectory instance to $p_k$, \ie, $\tau^{E,k}=\operatorname*{argmin}_{\tau^E\in\mathcal{T}} || p_k - f(\tau^{E}) ||^2_2 $ and $\mathcal{T}$ is the set of expert trajectories.


\subsection{Imitation Learning over Prototypes}
$\pi(a|\tau_t, \mathcal{P})$ is the policy layer that learns to take an action aligning with the prototypes from the experts' policy:
\begin{align}
    \pi(a|\tau_t, \mathcal{P}) = \phi([sim(f(\tau_t),p_1),..,sim(f(\tau_t),p_K)])
    \label{eq:pi}
\end{align}
where $\mathcal{P}= \{p_1,...,p_K\}$ is the set of prototypes embeddings, $\phi$ is a linear layer with sigmoid activation, and $sim(f(\tau_t), p_k) = -\|f(\tau_t)- p_k\|_2^2$ is the negative Euclidean distance measuring the similarity between the embedded vector and the prototype embedding $p_k$. The less similarity with more Euclidean distance.

We consider prototypical imitation learning with two base imitation learning models to learn the policy, \ie, behavior cloning (BC) and adversarial imitation learning (AIL).

The goal of BC is to mimic the action of the expert at each time step via supervised learning. 
{\small
\begin{align}
       \mathcal{L}_{IM_{BC}}   =\sum_{\tau^E\in\mathcal{T}} \sum_{(s^E,a^E)\sim\tau^E}[{\pi^E}(a^E|\tau^E_t)\log \pi(a|\tau_t, \mathcal{P})],
       \label{eqn:bcloss}
\end{align}}
where $\tau^E$ is the expert's trajectory, $\pi^E$ is the expert policy. 

On the other hand, the goal of AIL is to minimize the JS divergence between the expert trajectory distribution and trajectory distribution generated by our policy. 
{\small
\begin{align}
  \mathcal{L}_{\textit{IM}_{\textit{AIL}}} = D_{\textit{KL}}
\left(\rho_{\pi}\|\frac{\rho_{\pi}+\rho_{\pi^E}}{2}\right) 
+D_{\textit{KL}}  \left(\rho_{\pi^E} \|\frac{\rho_{\pi}+\rho_{\pi^E}}{2}\right),
\end{align}}
where $\rho_{\pi}$ and $\rho_{\pi^E}$ are discounted occupancy measures of our policy $\pi$ and the expert policy $\pi^E$.


In summary, the \textbf{full loss function} we are minimizing is:
{\small
\begin{align}
\mathcal{L}_{Full} =  w_1\cdot \mathcal{L}_{rep} +  w_2\cdot\mathcal{L}_{div} + w_3\cdot\mathcal{L}_{int}+ w_4\cdot\mathcal{L}_{IM_{loss}}
\label{eqn:full-loss}
\end{align}}
where $\mathcal{L}_{IM_{loss}}$ is the imitation learning loss with either BC or AIL, and $w_1, w_2, w_3, w_4, \in [0,1]$ are hyper-parameters to balance the weights of the three kinds of loss. We conduct grid-search to determine the value of these hyper-parameters (shown in Appendix).

\subsection{Reinterpretation as a Quadratic Model}
\label{sec:quadratic}
The policy $\pi$ is equivalent to a quadratic model with particular parameterization. By plugging in the similarity function, Equation~\ref{eq:pi} can be re-written as,
  \begin{equation}
  \pi  = -b_1\|f(\tau_t)- p_1\|_2^2  -... -b_K\|f(\tau_t)- p_K\|_2^2
 \end{equation}
where $b_k, k=1,...,K$ are the values of the linear neurons in the fully connected layer.
 By checking each term in $\pi$, we can see its linear form:
 {\small
 \begin{equation}
 \label{eqn:quadratic}
     -b_k\|f(\tau_t)- p_k\|_2^2  = -b_kf(\tau_t)^Tf(\tau_t) + 2b_kp_k^Tf(\tau_t) - b_kp_k^Tp_k
 \end{equation}}
where the first term is a quadratic term with regard to $f(\tau_t)$, the second term can be treated as $w_k^Tf(\tau_t)$ where $w_k = 2b_k p_k$ and the final term can be treated as a constant term with regard to $f(\tau_t)$.

Starting from the above observation, we can treat the action as a summation of $K$ quadratic functions with the same sign in quadratic coefficients with regard to $f(\tau_t)$, which means the relationship between the action and $f(\tau_t)$ can be decomposed to at most two pieces and within each piece the relationship is monotonic. This observation makes our learned policy easier to interpret.

\subsection{Unified human-in-the-loop learning}
Systematic noise in trajectories leads 
to sub-optimal prototype embedding, prototype selection, and final policy. Our efficient HITL module exploits active feedback to refine the learned ``policy over prototypes''. Next, we describe the two aspects of this module \textemdash~ (1) The query framework to elicit relevant active human feedback (2) Incorporating knowledge from such feedback into the learning process. Human feedback can be of two forms: (1) feedback about the quality of prototype embedding, prototype alignment and diversity, (2) feedback about the risk of predicted actions. 

\subsubsection{Query Framework (knowledge elicitation):} 
Posing focused queries at relevant points is essential for getting useful and relevant knowledge. In a traditional active learning setting, just prediction/label uncertainty is a sufficient heuristic to identify relevant query points. However, in our context, we need richer feedback on both relevant prototypes and output actions. 
Thus queries at step $t$ is a set of tuples $ \mathcal{Q}_t=\{\langle p_{q_{t}}, a_{q_t}\rangle | p_{q_{t}}\in\mathcal{P}_{q_t}, a_{q_t}\in\mathcal{A}_{q_t}\}$ that comprises a set of prototypes embeddings $\mathcal{P}_{q_t} \subseteq \mathcal{P}$ and predicted action(s) $\mathcal{A}_{q_{(t)}}$ that need feedback. $\mathcal{Q}_{(t)}=\emptyset$ signifies there is no query at that step. Now $\mathcal{P}_{q_{t}} = \mathcal{P}_\mu \cap \mathcal{P}_d$ where $\mathcal{P}_\mu$ is defined as the set of uncertain prototypes,
\begin{align}
\nonumber & \mathcal{P}_\mu = \{p_k \in \mathcal{P}| \mu(p_k) \geq \mathbb{U}_p \}\\
&  \mu(p_k) = \mathbb{H}(P(||f(\tau)-p_k||_2^2)), \tau \in D_k
\end{align}
where $\mathbb{U}_p$ is the uncertainty threshold and uncertainty is the prototype cluster entropy $\mathbb{H}$ via the distribution $P(\cdot)$ induced over L2 distances defined earlier. $\mathcal{P}_d$ comprises prototype embeddings with `top-N' average distance of the form, 
\begin{align}
  \nonumber \mathcal{P}_d = \arg\max_{\mathbf{p}\subset \mathcal{P}}\left(\frac{1}{|D_k|}\sum_{\tau\in D_k }\overline{||f(\tau)-p_k||_2^2}\right), \left|\mathbf{p}\right|=N
\end{align}
where $\mathbf{p} \models p_k$ is the top-N set. Now, $a_{q_{t}}$ is identified via both prediction uncertainty $\mu(a_{q_{t}}) = \mathbb{H}(\pi(a|\tau_t,\mathcal{P})) \geq \mathbb{U}_a$ ($\mathbb{U}_a$ is the action uncertainty threshold), and oversubscription risk which indicates if output action is less than the expert action $a < a^E$. Note that lower oversubscription action/rate means higher risk (cf. Section~\textit{Domains(Datasets)}). 
\emph{This query formulation is aligned with the interpretable quadratic form of action/policy (\cref{sec:quadratic}). The square term trajectory embeddings in \cref{eqn:quadratic} can lead to higher risk in case of label noise. Hence action feedback is paramount. Whereas, the feedback on prototype $p_k$ is essential to control the coefficient of the second term $2b_kp_k^Tf(\tau_t)$}

\subsubsection{Feedback representation:} Given $\mathcal{Q}_{t}$, the expert can provide feedback on prototype quality and membership/diversity or on action predictions. Feedback on prototype quality at step $t$ for a prototype embedding $p_j$ is an up/down vote ($\mathfrak{F}(p_{k}|t)=\uparrow/\downarrow = +1/-1$). Then we use cumulative feedback (over previous iterations) $\mathfrak{F}(p_k) = \sum_{t'=1}^t \mathfrak{F}(p_k|t')$. We obtain action feedback $\mathfrak{F}(a)$ in a similar fashion. Feedback on prototype diversity is complicated. For a given pair $p_i,p_j$, expert may choose to up/down-vote the pair or merge $p_i, p_j$ or split them further. \textbf{(1)} If the prototypes are merged, it gets the mean of the embedding vectors of $\vec{p_k} \gets \frac{1}{2}(\vec{p_i}\bigoplus\vec{p_j})$, \textbf{(2)} if a prototype  $p_k$ is split, the new prototype is initialized with embedding $f(\tau): \max_{\tau \in D_k}||p_k - f(\tau) ||^2_2$ and retrained.

\subsubsection{Prototype/policy refinement:} In our context it is difficult to design either inductive bias distribution or proper constraints from feedback. So we allow the feedback to control and scale the loss via exponential `advice potential gates' (cf. K-CLN \cite{das2021human}). Advice potentials selectively alter the loss such that training moves in the direction that is expected to produce better model parameters. 
\begin{definition}[Advice potential gate $\Phi$]
\label{def: advice_gate}
An advice potential gate is a product term of an exponential form $\Phi(x) = e^{-\mathcal{G}(\mathfrak{F}(x))}$ where $-\infty \leq \mathfrak{F} \leq +\infty$ is the cumulative feedback. $\mathcal{G}$ scales the unbounded (cumulative) $\mathfrak{F}$ to $[-1,+1]$
\end{definition}
With the available feedback over a prototype ($\mathfrak{F}(p_k)$), action ($\mathfrak{F}(a)$) or between prototypes ($\mathfrak{F}(p_i,p_j)$) we modify the loss function of \ourmethod in Equation~\ref{eqn:full-loss} as,
\begin{align}
    \nonumber \mathcal{L}'_{Full} = & w_1\cdot \mathcal{L}_{rep}\times\Phi(p_k) +  w_2\cdot\mathcal{L}_{div} \times \Phi(p_i,p_j) \\
    & + w_3\cdot\mathcal{L}_{int} \times \Phi(p_k) + w_4\cdot\mathcal{L}_{IM_{loss}}\times \Phi(a),
\end{align}
where operator $\times$ takes effect on related prototypes/actions when computing loss. Advice potentials upscale or downscale the relevant loss components as per cumulative feedback 
and adaptively control the degree of loss for relevant prototypes/actions, navigating the loss landscape towards better models. 

\textbf{HITL contributions:} 
Our HITL component makes 3 primary contributions.
Firstly, unlike existing  human-in-the-loop IL approaches, where human feedback is limited to selecting or re-prioritizing actions, \ourmethod allows humans to influence the selection and learning of prototypes that represent diverse oversubscription policies (Figure~\ref{fig:model}). 
Secondly, since prototypes essentially project decision space into a different lower dimensional manifold, the HITL feedback over prototype quality and diversity contains much richer knowledge than action selection. Finally, and most importantly \ourmethod elicits active feedback from the human-in-loop by identifying the prototypes or predicted actions that need more information/help (knowing-what-it-knows), making our HITL component highly efficient. 
\section{Experiments}
We empirically evaluate our method on virtual CPU oversubscription scenario in Microsoft 365's internal (1$^{st}$ party) cloud platform. Additionally, to highlight the generality of \ourmethod we also evaluate on a novel airline ticket overbooking domain described later. We collect real data from these domains and propose respective simulators for evaluation that are described next.  

\subsection{Domains (Datasets)}\label{sec:domain}
\textbf{Virtual CPU oversubscription.} A Virtual Machine (VM) is a virtualized instance of a computer that runs on a physical server and accesses computing resources to perform functions~\cite{li2010survey}. Users purchase VMs from cloud platform providers to host applications and services. The cloud platform contains many of physical servers, \ie, \textit{Nodes}, and each node hosts a certain number of VMs. 

CPU bottlenecks are more severe and common than memory and network in cloud~\cite{mahapatra1999processor}. As shown in Figure~\ref{fig:vcpuoversub}, three VMs are placed in the same node, and the CPU dimension is fully packed with unused memory, \ie, stranded memory displayed as the grey box. If oversubscribing actual physical CPU size, additional VMs can be allocated on this node reducing stranded memory. Also, virtual CPU (vCPU) represents a portion or share of the underlying physical CPU that is assigned to a VM. Thus vCPU is the sellable billing unit in cloud platforms. So, we focus on vCPU oversubscription. 

We collect real data from Microsoft's cloud platform 
for internal users, \ie, responsible owners of M company services and applications. Such internal users host their services on VMs whose vCPU usage demonstrates patterns. For example, services like emails and work-related software demonstrate diurnal and weekly patterns in regions with similar time zones 
corresponding to work and social activities, resulting in the most requests and traffic income on daytime and weekdays. Thus the vCPU usage is high during these time periods while low at nighttime and weekends. On the other hand, services providing social media and gaming applications show different vCPU usage patterns when peak/hot usage happens in spare time. Moreover, other non-user-facing services running regularly, like monitoring and maintenance services, do not show daily or weekly patterns but display patterns caused by underlying configurations from service teams. The diverse vCPU usage patterns of services motivate us to adaptively oversubscribe vCPUs of VMs from each service. For example, oversubscribe services with low vCPU usage, given the context. 

We collect two-week data of VM features, including the usage of vCPU, memory, and network, that belong to 30 randomly sampled services. As the vCPU usage has a lot of fine-grained variances, we take the peak usage in the one-hour bucket as the representative data point, making oversubscription conservative. Note that in Figure~\ref{fig:vcpuoversub}, VMs are allocated onto nodes, and VMs from different services can be collocated in the same node. Then, we propose a simplified allocation simulator that allocates VMs via Best-Fit allocation policy~\cite{hadary2020protean}. VMs are allocated after vCPU oversubscription.



\noindent\textbf{Other domains - Airline ticket overbooking:} 
As noted earlier to highlight how our approach can seamlessly work on any domain to optimize adaptive oversubscription, based on utilization patterns we present Airline ticket overbooking scenario. Flight overbooking, \ie, selling more tickets than the available seats, is a common practice that allows airline companies to improve their load factors and increase revenues~\cite{nazifi2021proactive}. Yet, the difficulty in estimating ticket demands and no-shows results in inappropriate overbooking strategies, such that users with tickets cannot onboard, \ie, offloaded. In general, flight ticket demands show quarter patterns that peak tourist seasons have higher flight demands~\cite{suryani2010air, banerjee2020passenger}. This motivates adaptive oversubscription of flight tickets.
We collect airline passengers' data from the overbooking reports of the U.S. Department of Transportation (DOT). The dataset covers overbooking information of 32 airline companies in the U.S. from 1998 to 2021~\footnote{https://www.bts.gov/denied-confirmed-space}, reported quarterly. Each quarter's data includes offloaded number of passengers (voluntary/involuntary) and the onboard number of passengers. As the overbooking rate of each airline company and the actual demands are not reported, we generate synthetic overbooking rate and flight demands to create a semi-synthetic dataset. We sample the overbooking rate within a range on 3\%-5\%, as per common industry practice~\cite{suzuki2006net}. With the popularity of electronic tickets, no-shows are less compared to the paper-ticket period~\cite{nazifi2021proactive}, we assume no-show rate to decay with years. Then the demand is the aggregation of bumped, no-shows, and onboard passengers. We developed a simulator trained with GBDT on the semi-synthetic data that outputs the overbooking rate of the next quarter given current quarter features. 


\subsection{Experiment Configurations}
All experiments are performed on Ubuntu 20.04 LTS system with Intel(R) Xeon(R) CPU E5-2690 v3 @ 2.60GHz CPU, 112 Gigabyte memory and single NVIDIA Tesla P100 accelerator. Detailed settings are listed below.

\subsubsection{Base Learner:}
We implemented our method by extending on top of Behavior Cloning as the base imitation learner in our experimental version. As explained later, we also use Behavior Cloning as one of the baselines.
We set the learning rate as $1e-2$, MLP unites is 64, batch size is 128, optimizer is Adam. We tune the number of prototypes in $\{2,3,4,5,6\}$ and set 3 for vCPUs and 4 for airplane tickets respectively with the optimal search value. We tune the weights of different loss components within $[0.1, 0.2, 0.3,...,1.0]$ via grid-search. Finally, we obtained the best value on $w_1=0.8$, $w_2=0.1$, $w_3=0.1$, $w_4=1.0$.

\subsubsection{State, Action, Reward}
In our vCPU oversubscription problem, the state $s_t$ consists of the feature of the historical CPU usage rate, users' memory, CPU, and network requests $q_t$ for each VM, Nodes' capacity and etc. at time step $t$. The action $a_t\in[0,1]$ indicates the oversubscription rate which indicates we will allocate a VM with $a_t * q_t$ memory, network, and CPU. The reward $r_t = -h_t + m_t$, where $h_t$ indicates the number of hot nodes and $m_t$ indicates the number of saved vCPUs or remain Core. As for the flight tickets oversubscription problem, the state $s_t$ indicates the historical sold tickets, the number of onboarding customers, the number of seats $e_t$ in the airplane and etc. The action $a_t\in[0,1]$ indicates the oversubscription rate which which we will sell $a_t*e_t$. The reward function $r_t = -c_t + o_t$ , where $c_t$ indicates the compensation cost and $o_t$ indicates the profits. In summary,  the state space is factored with a hybrid feature vector, including temporal features. The action/decision space is continuous, \ie, the oversubscription level. This makes our problem fairly complex to be solved in a straightforward behavior cloning fashion. Instead, \ourmethod embeds into a latent space of equivalence classes or prototypes and exploiting approximate symmetries among the trajectory patterns. 

\subsubsection{Configuration settings of HITL:} Some of the settings of the HITL module as follows,

\code{use\_hitl} flag allows us to enable disable the human-in-the-loop feedback system. We set is \code{True} in our \textsc{ProtoHAIL} experiments.
    
\code{FREQUENCY} parameter in the feedback object of \textsc{ProtoHAIL} controls the frequency of queries to the human. The default is set to $10$ iterations. Note that this does not necessarily mean that a query will definitely be generated at every $100$ iterations. Queries are subject to the query generation formulation given the quality of prototypes and predictions as outlined in the manuscript. However, this is a lower bound. If more queries are generated in between 2 consecutive 100 iterations then they will be skipped. 
    
\code{Uncertainty Threshold} $Tr$, which is defaulted to 0.8 in our experiments, post empirical observation. 

Other hyperparameters are inherited from the base prototypical imitation learning module. 
HITL experiments on vCPU domain has been performed with a \code{batch\_size = 128}, which also the default in base protoypical IL. In case of airplane ticket overbooking, we performed HITL experiments with \code{batch\_size = \{32, 64, 128\}}, all of them giving us the same results that have been reported in the paper. Base IL uses are batch size of $1$ due to the way the training loop has been designed for this domain, however in HITL batch size of $1$ does not allow proper prototype uncertainty computations so we experimented with batch sizes $>1$ and we made sure the performance is not affected. 
Another important aspect is the number of epochs/iterations are same as default values in based code IL (for example 300 in case of vCPU experiments). However, if a \code{split} or \code{merge} feedback is given and the prototype structure is altered we introduce $30$ shadow training iterations after such feedback for the change to take effect from the next explicit iteration. So the effective iterations may be slightly higher; $total~iterations = base~iterations + \#(merge/split)\times 30$.

\subsection{Experimental Settings}
We evaluate our approach on the vCPU oversubscription as well as the ticket overbooking domains, focusing on the interpretability and effectiveness from three aspects: 
(1) learned prototypes visualization: We visualize prototypes to analyze the policy learned by our method. 
(2) Task inference. We show how \ourmethod learns to compose the prototypical options to solve both continuous and discrete control tasks.
(3) The performance of \ourmethod in these tasks. In particular, we outline two evaluation metrics for vCPU oversubscription task based on how real cloud platforms behave, \ie, \textit{hot node} and \textit{remain core}. Specifically, a node hosts several VMs (refer to Section~\ref{sec:domain}) and reserves 15\% resources as a buffer. If the node CPU utilization is higher than 85\%, the node is a \textit{hot node}~\cite{qiao2021intelligent} which diminishes the performance of services running on it (risk). The other evaluation metric, \textit{remain core}, measures the benefits of oversubscription. It is defined as the remaining cores that could be potentially used by additional VMs after vCPU oversubscription (benefit). Adapting this to the flight ticket overbooking task, \textit{compensation cost} refers to the monetary compensations for offloaded passengers (risk), and \textit{profit} refers to the extra revenues gained by overbooking (benefit).
Fewer hot nodes/ lower compensation costs with more remaining cores/ more profits on vCPU and airline oversubscription resp., as illustrated in Table~\ref{tab:compare}, highlight how \ourmethod outperforms baselines. 

Fewer hot nodes/ lower compensation costs with more remaining cores/ more profits on vCPU and airline oversubscription resp., as illustrated in Table~\ref{tab:compare}, highlight how \ourmethod outperforms baselines. Our detailed ablation studies presented in \cref{sec:ablation} on the learned prototypes and the policies as well as the HITL module highlights how, (1) earned prototypes are interpretable representatives of an equivalence class of trajectories and (2) efficient HITL refinement leads to better prototypes. Further studies and 
additional results on hyper-parameter tuning and pressure tests are presented later in \cref{sec:hypertuning,sec:pressure}.







\subsection{Comparisons against SOTA Baselines}\label{sec:competitors}
\label{baseline}
\textbf{Baselines:} We compared three baselines, (1) \textbf{Heuristic policy}: Grid-search, which searches for the overall oversubscription rate to improve the benefits by reducing the costs/risk. Moving average, which averages the historical usage rate as the current time step's oversubscription action. (2) \textbf{Imitation learning}: BC~\cite{pomerleau1991efficient}, GAIL~\cite{ho2016generative} and human-guided IL such as Dagger (with 20 steps of human guidance). (3) \textbf{Reinforcement learning}: DDPG \cite{lillicrap2016continuous} for continuous action space.

\begin{table}[t] 
  \centering
  \caption{Results of different learning strategies on {\em vCPU oversubscription} and {\em flight tickets overbooking}.}
  \begin{tabular}{
    l@{}c@{\hspace{6pt}}c@{\hspace{4pt}}c@{\hspace{6pt}}c
    }
   \toprule
     \multirow{2}{*}[-2pt]{Approach} & \multicolumn{2}{c}{\text{{\em vCPU Oversub.}}} & \multicolumn{2}{c}{ \text{{\em Flight Tickets}}} \\
    \cmidrule(r{2pt}){2-3} \cmidrule(l{2pt}){4-5} 
      & \text{Hot Node/Risk} & \text{Core} & \text{Cost} & \text{Profit} \\
    \midrule
    {\em Grid-search} & 0\% & 7450 & 0M & 0M \\ 
     {\em Moving Average} & 1.39\% & 7628 & 0.96M & 6.79M \\
      {\em DDPG} & 1.47\% & 5030 & 12.37M & 2.35M \\
      {\em Behavior Cloning} & 1.19\% & 7870 & 1.47M & 7.21M \\
      {\em GAIL} & 1.2\% & 6980 & 2.74M & 4.56M \\
      {{\em Dagger} ({\em \scriptsize 20 time steps})} & 0.96\%& 7938 & 0.47M & 6.95M \\
       {\ourmethod ({\em \scriptsize w/o HITL})} & 0\% & 8153 & 0.31M & 8.79M \\
       {\ourmethod} & 0\% & 8161 & 0.14M & 13.65M \\
    \bottomrule
  \end{tabular}
    \label{tab:compare}
    \vspace{-10pt}
\end{table}

\textbf{Results:} 
We run five seeds to conduct Wilcoxon signed-rank tests to check the statistical significance of the results. It shows that all the p-values of Wilcoxon signed-rank tests at 95\% confidence level are smaller than 0.05.
We further observe that: (1) DDPG (RL) fails in both scenarios as it is hard to optimize two competing objectives simultaneously in  standard RL. (2) The heuristic policies, \ie, Grid-search and Moving Average, show robust performance on vCPU oversubscription. But Grid-search fails on airline overbooking due to the high risk of that problem, 
\ie, the ticket sales are highly dynamic, and a static oversubscription rate fails to adapt to different situations. (3) BC beats GAIL, possibly because BC has better sample efficiency 
and, in oversubscription scenarios, training samples are limited. Furthermore, the human-guided approach (Dagger) shows the second-best performance, demonstrating the benefits of human guidance in this domain. 
We set 20 steps of human guidance in Dagger, which is fair enough because it accounts for almost 1/6 of the trajectory length on average. Also, in our model, the number of queries to humans in HITL is much smaller than 20 (\eg, 6 queries on average in the vCPU experiments and 8 queries on average in the Air Ticketing experiments). 
(4) \ourmethod outperforms all, both with and without HITL, as it captures different classes of patterns via learning a set of prototypes, and HITL refinement of such classes further improves the performance (Core/Profit) and mitigates risk (Hot node/ ost). Although grid-search, the most conservative policy, is at par in terms of risk mitigation, \ourmethod achieves both the least risk and highest benefit in both domains.

\subsection{Ablation Studies}
\label{sec:ablation}
\begin{figure}[htbp]
    \centering
    \subfigure[Prototypes in vCPUs oversubscription.]{
    \includegraphics[width=.40\textwidth]{ 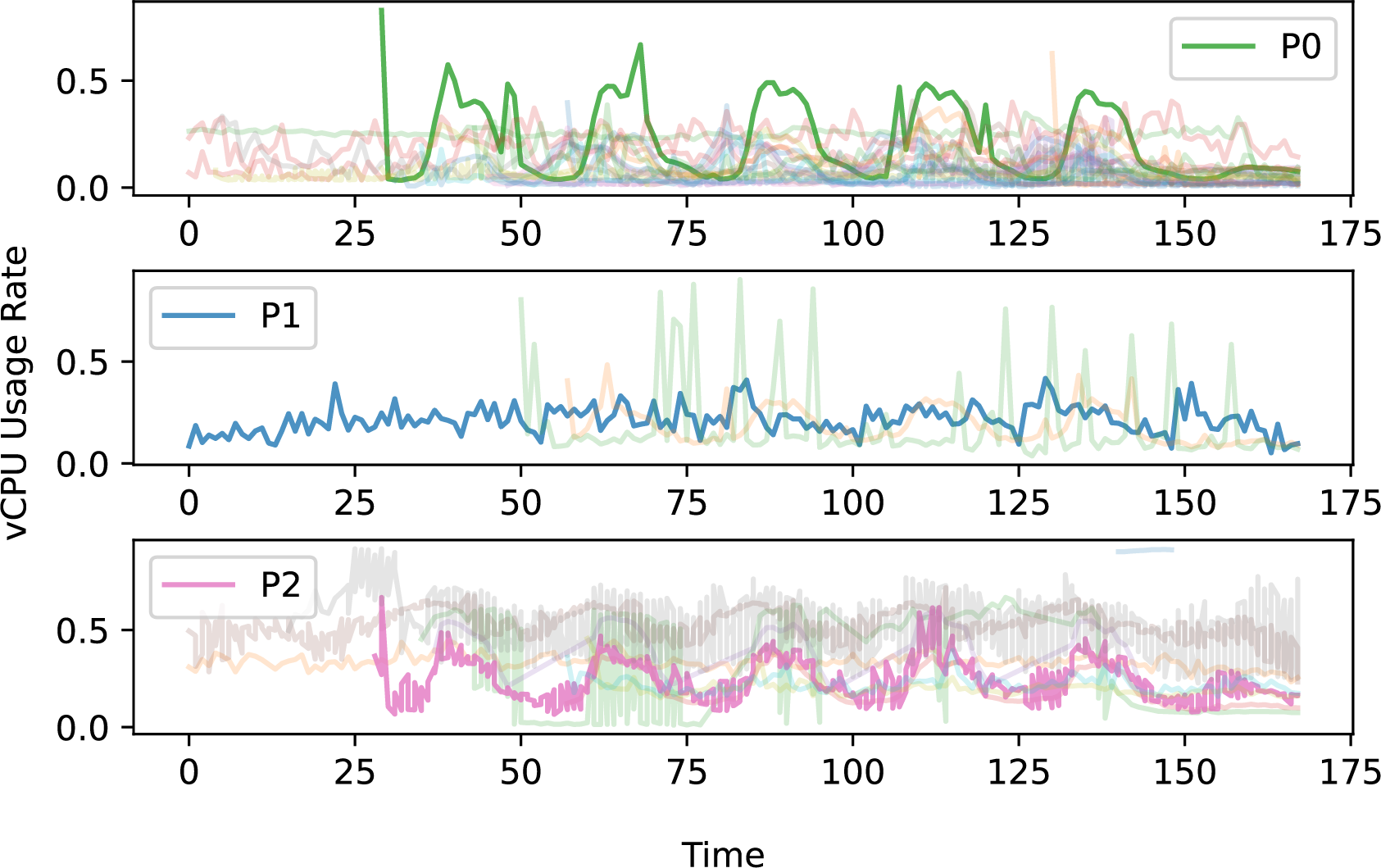}
    \label{fig:vCPU}
    }
    \subfigure[Prototypes in airplane tickets oversubscription.]{
    \includegraphics[width=.40\textwidth]{ 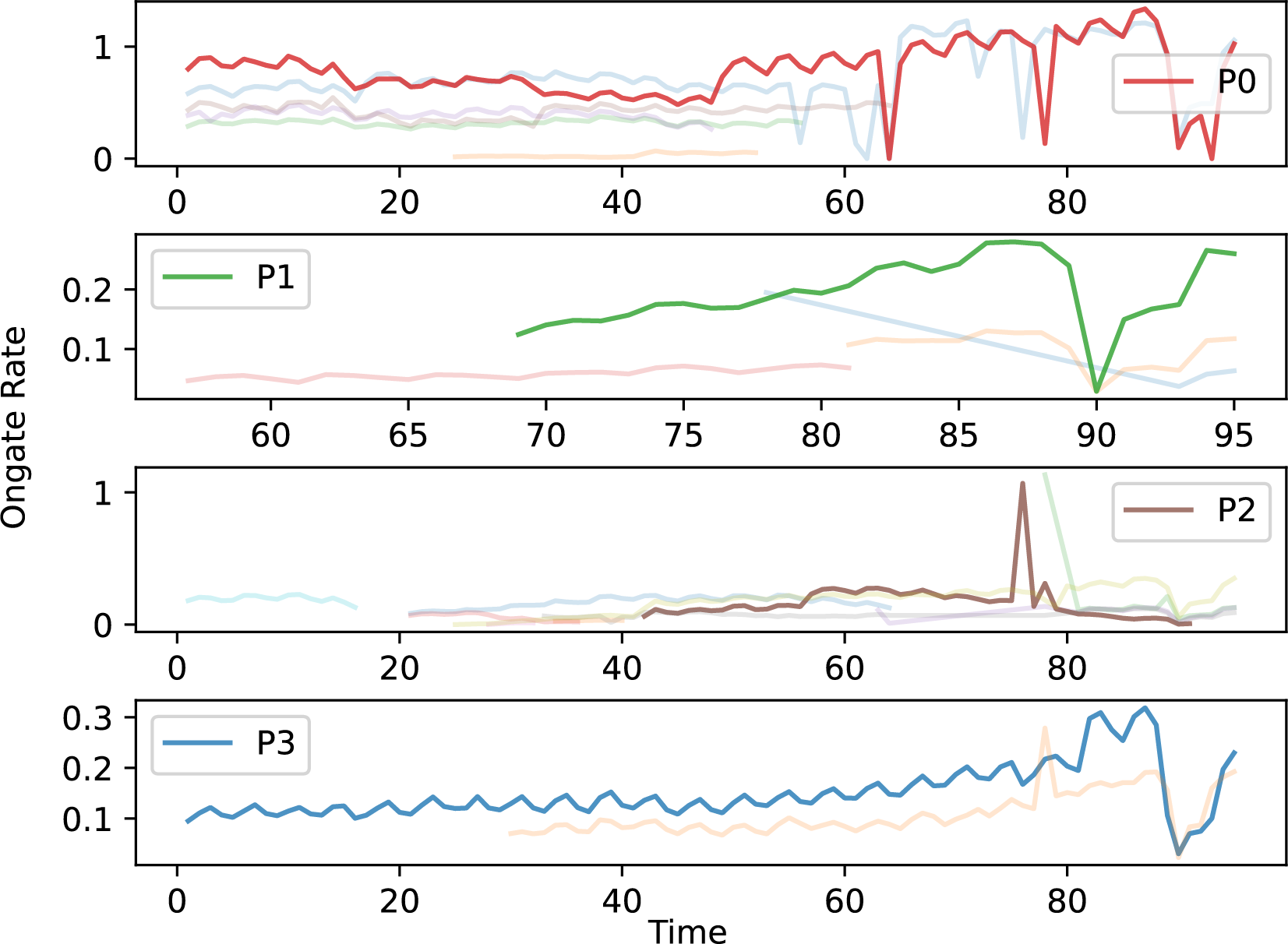}
    \label{fig:airplane}
    }
    \caption{Learned Prototypes with \ourmethod}
    \vspace{-4mm}
\end{figure}

\subsubsection{Analysis of the Learned Prototypes:} We learn $3$ and $4$ prototypes in vCPU oversubscription and airline overbooking, respectively. \ourmethod learns these prototypes and represents them via finding the nearest trajectories.
In Figure~\ref{fig:vCPU}, different sub-figures show the trajectories in different prototype clusters. The prototype embedding is highlighted, and the trajectories are in transparent color. We analyze the interpretability of the policy via the learned prototypes: 
prototypes with hourly patterns are email/work-related services ($P0$), non-user-facing services that do not show temporal patterns ($P1$), and social media/gaming services ($P2$).
This demonstrates the capability of \ourmethod in learning interpretable prototypes.

As for the airplane tickets (Figure~\ref{fig:airplane}), \ourmethod learns $4$ prototypes with different trends. $P0$ indicates policies within a stable constant range. $P1$ indicates continuously increasing levels with occasional drops. $P2$ indicates policies that would be sensitive to the environment and quickly increase action value. $P3$ indicates a smooth increase over time. 

\textbf{Limitation}: Airline ticket overbooking data does not show a clear pattern due to the drop in demands during COVID-19~\cite{amankwah2021covid}.

\subsubsection{Analysis of human in the loop:}
As Table~\ref{tab:compare} clearly illustrates, our HITL module contributes to additional risk-minimization (and often higher benefits) in both domains with minimal queries to the human expert. There are two aspects to efficient HITL, (1) Querying only at relevant stages and (2) Effective refinement. In vCPU oversubscription, \textit{we initialize with 6 prototypes, based on prior insights, which works reasonably well during the early stages of training.} However, once the learning plateaus we observe, (1) for services mostly in prototype $P1$, the predicted oversubscription rate remains significantly higher than usage (2) in $P5$, predicted oversubscription is less than usage, \textit{an overloading risk} (3) $P4$ has almost no services in the cluster, making it marginally redundant. \ourmethod automatically detects such nuances and poses targeted queries, at only particular stages, to actively obtain feedback, similar to the example shown below.\\
\noindent \fbox{
\parbox{0.97\columnwidth}{
{\scriptsize
STEP 110; Query: P=[5, 1]=[0.9904377 1.0770873] seem unstable. 
Please suggest: (1=good, 0=none, -1=bad OR "split $<$prototype\#$>$"). 
\textit{\$ input: 1 = -1} \\
STEP 240; Following prototype pairs seem redundant [(4,1), (2,0), (2,5)]. Wish to merge any 
(merge $<$prototype\#$>$)? \textit{\$ input: merge 4 1} 
}}}

It automatically identifies $P5/P1$ as problematic as well as P4 being redundant or sharing characteristics with P1; probes about possible merging. Feedback on predicted actions with overloading risk is elicited in a similar fashion. In our experiments, the total number of queries required to reach reported stable performance was $\leq 10$ demonstrating how efficient our HITL is.
The feedback is then used to jointly refine both the prototype memberships and the embedding function. Figure~\ref{fig:vCPU}/\ref{fig:airplane} shows how the number of prototypes has reduced to 3 and 4 post-refinement. 
Additionally, the queries may be sensitive to hyperparameters such as thresholds($\mathbb{U}_p$ \& $\mathbb{U}_a$). So we performed a grid search and observed that any threshold in a range of $(0.5,0.75)$ led to a stable performance on average. Reported results are with 0.55 as thresholds.

\subsection{Additional studies}
\subsubsection{Hyper-parameter Tuning: }
\label{sec:hypertuning}
The number of prototypes is the key hyper-parameters for our model.
We tune the number of prototypes in $\{2,3,4,5,6\}$ and set 3 for vCPUs and 4 for airplane tickets respectively with the optimal search value.  
The detailed results are shown in Table~\ref{tab:numberParameter}.
\begin{table}[htbp] 
  \centering
  \caption{Hyper-parameter tuning.}
  \begin{tabular}{
    l@{}c@{\hspace{6pt}}c@{\hspace{4pt}}c@{\hspace{6pt}}c
    }
   \toprule
     \multirow{2}{*}[-2pt]{\# options} & \multicolumn{2}{c}{\text{{\em vCPU Oversub.}}} & \multicolumn{2}{c}{ \text{{\em Flight Tickets}}} \\
    \cmidrule(r{2pt}){2-3} \cmidrule(l{2pt}){4-5} 
      & \text{Hot Node} & \text{Core} & \text{Cost} & \text{Profit} \\
    \midrule
    {\em 2} & 0.26\% & 8067 & 0.32M & 12.59M \\ 
    {\em 3} & 0\% & 8161 & 0.27M & 13.10M \\
     {\em 4} & 0\% & 8092 & 0.14M & 13.65M \\
      {\em 5} & 0.05\% & 8154 & 0.22M & 13.15M \\
      {\em 6} & 0.07\% & 8150 & 0.16M & 12.64M\\
    \bottomrule
  \end{tabular}
    \label{tab:numberParameter}
\end{table}

There are 2 primary hyperparameters in the HITL module; the uncertainty thresholds $\mathbb{U}_p, \mathbb{U}_a$. While they do not directly affect the loss surface, they control query generation. Extremely conservative (high) uncertainty thresholds lead to extremely focused queries and reduce the number of queries. However, it often leads to missing out on prototypes that really do need feedback, since the final query is an intersection with top-N prototype embeddings. On the other hand, the relaxed (lower) thresholds lead to too many queries. Another minor hyperparameter is `N', the cardinality of top-N choices of in the case of $p_d$. All results are reported with $N=5$. However, we did perform a grid search over all 3 hyperparameters and observed that for thresholds a range of $(0.5,0.75)$ was mostly appropriate, with 0.55 being the value used for reported results. For N higher values make no sense since $p_d$ intersects with $p_\mu$ anyway. Values of N lower than 3 resulted in $p_\mu \cap p_d = \emptyset$.

\subsubsection{{Pressure Test: }}
\label{sec:pressure}
Our experiments are based on real-world datasets. It is impractical to conduct pressure tests from two aspects: a) we cannot do endless vCPU oversubscription if there is no stranded memory available to place additional VMs. b) the real-world CPU utilization distribution would rarely make many hot nodes as their CPU utilization will not peak simultaneously. Nevertheless, we also experimented on the high-pressure test by manually reducing the hot ratio (which is much smaller than the real-world value), the results are shown as follows, where ProtoHAIL also achieves the best performance on both metrics compared with the best two baselines.
\begin{table}[htbp]
    \centering
    \caption{Pressure Test}
    \begin{tabular}{ccc}\toprule
         Method & Hot Node & Remain Core\\
         \midrule
         Behavior Cloning & 92.61\% & 28\\
         Dagger & 90.74\% &23\\
         ProtoHAIL& 86.25\%	 & 34\\
         \bottomrule
    \end{tabular}
    \label{tab:class_setting}
\end{table}

\section{Conclusion}
\label{conclusion}
We presented our novel prototypical imitation learning framework with the human-guided refinement that addresses a critical decision-making problem of optimal oversubscription with competing objectives of efficient resource utilization with minimal risk. We show via extensive and ablation studies on real domains how learning representative prototypes not only achieves effective oversubscription policies but also discovers implicitly interpretable classes of utilization patterns. Such interpretability, in turn, allows for efficient active human-in-the-loop prototypes and policy refinement that provides further benefits and risk hedging. Through extensive empirical evaluation and case studies we have exhibited how our framework is extremely effective on Microsoft's internal cloud platform and can be generalized seamlessly to other oversubscription domains as well. In future work, we plan to scale and deploy our framework on actual oversubscription systems or even extend \ourmethod to other problems with similar properties. Combining our prototypical IL with a continual learning RL framework is also an interesting future direction. Finally, provable regret bounds with human-in-the-loop policy refinement is an open problem and are even more complex with prototypes.

\newpage
\bibliographystyle{ACM-Reference-Format}
\bibliography{sample-base}


\begin{thebibliography}{43}


\ifx \showCODEN    \undefined \def \showCODEN     #1{\unskip}     \fi
\ifx \showDOI      \undefined \def \showDOI       #1{#1}\fi
\ifx \showISBNx    \undefined \def \showISBNx     #1{\unskip}     \fi
\ifx \showISBNxiii \undefined \def \showISBNxiii  #1{\unskip}     \fi
\ifx \showISSN     \undefined \def \showISSN      #1{\unskip}     \fi
\ifx \showLCCN     \undefined \def \showLCCN      #1{\unskip}     \fi
\ifx \shownote     \undefined \def \shownote      #1{#1}          \fi
\ifx \showarticletitle \undefined \def \showarticletitle #1{#1}   \fi
\ifx \showURL      \undefined \def \showURL       {\relax}        \fi
\providecommand\bibfield[2]{#2}
\providecommand\bibinfo[2]{#2}
\providecommand\natexlab[1]{#1}
\providecommand\showeprint[2][]{arXiv:#2}

\bibitem[Achiam et~al\mbox{.}(2017)]%
        {achiam2017constrained}
\bibfield{author}{\bibinfo{person}{Joshua Achiam}, \bibinfo{person}{David Held}, \bibinfo{person}{Aviv Tamar}, {and} \bibinfo{person}{Pieter Abbeel}.} \bibinfo{year}{2017}\natexlab{}.
\newblock \showarticletitle{Constrained policy optimization}. In \bibinfo{booktitle}{\emph{International conference on machine learning}}. PMLR, \bibinfo{pages}{22--31}.
\newblock


\bibitem[Amankwah-Amoah(2021)]%
        {amankwah2021covid}
\bibfield{author}{\bibinfo{person}{Joseph Amankwah-Amoah}.} \bibinfo{year}{2021}\natexlab{}.
\newblock \showarticletitle{COVID-19 pandemic and innovation activities in the global airline industry: A review}.
\newblock \bibinfo{journal}{\emph{Environment International}}  \bibinfo{volume}{156} (\bibinfo{year}{2021}), \bibinfo{pages}{106719}.
\newblock


\bibitem[Banerjee et~al\mbox{.}(2020)]%
        {banerjee2020passenger}
\bibfield{author}{\bibinfo{person}{Nilabhra Banerjee}, \bibinfo{person}{Alec Morton}, {and} \bibinfo{person}{Kerem Akartunal{\i}}.} \bibinfo{year}{2020}\natexlab{}.
\newblock \showarticletitle{Passenger demand forecasting in scheduled transportation}.
\newblock \bibinfo{journal}{\emph{European Journal of Operational Research}} \bibinfo{volume}{286}, \bibinfo{number}{3} (\bibinfo{year}{2020}), \bibinfo{pages}{797--810}.
\newblock


\bibitem[Baset et~al\mbox{.}(2012)]%
        {baset2012towards}
\bibfield{author}{\bibinfo{person}{Salman~A Baset}, \bibinfo{person}{Long Wang}, {and} \bibinfo{person}{Chunqiang Tang}.} \bibinfo{year}{2012}\natexlab{}.
\newblock \showarticletitle{Towards an understanding of oversubscription in cloud}. In \bibinfo{booktitle}{\emph{2nd USENIX Workshop on Hot Topics in Management of Internet, Cloud, and Enterprise Networks and Services (Hot-ICE 12)}}.
\newblock


\bibitem[Breitgand and Epstein(2012)]%
        {breitgand2012improving}
\bibfield{author}{\bibinfo{person}{David Breitgand} {and} \bibinfo{person}{Amir Epstein}.} \bibinfo{year}{2012}\natexlab{}.
\newblock \showarticletitle{Improving consolidation of virtual machines with risk-aware bandwidth oversubscription in compute clouds}. In \bibinfo{booktitle}{\emph{2012 Proceedings IEEE INFOCOM}}. IEEE, \bibinfo{pages}{2861--2865}.
\newblock


\bibitem[Brown et~al\mbox{.}(2018)]%
        {pmlr-v87-brown18a}
\bibfield{author}{\bibinfo{person}{Daniel~S. Brown}, \bibinfo{person}{Yuchen Cui}, {and} \bibinfo{person}{Scott Niekum}.} \bibinfo{year}{2018}\natexlab{}.
\newblock \showarticletitle{Risk-Aware Active Inverse Reinforcement Learning}. In \bibinfo{booktitle}{\emph{Proceedings of The 2nd Conference on Robot Learning}}.
\newblock


\bibitem[Das et~al\mbox{.}(2021)]%
        {das2021human}
\bibfield{author}{\bibinfo{person}{Mayukh Das}, \bibinfo{person}{Devendra~Singh Dhami}, \bibinfo{person}{Yang Yu}, \bibinfo{person}{Gautam Kunapuli}, {and} \bibinfo{person}{Sriraam Natarajan}.} \bibinfo{year}{2021}\natexlab{}.
\newblock \showarticletitle{Human-Guided Learning of Column Networks: Knowledge Injection for Relational Deep Learning}. In \bibinfo{booktitle}{\emph{COMAD/CODS}}.
\newblock


\bibitem[De~Haan et~al\mbox{.}(2019)]%
        {de2019causal}
\bibfield{author}{\bibinfo{person}{Pim De~Haan}, \bibinfo{person}{Dinesh Jayaraman}, {and} \bibinfo{person}{Sergey Levine}.} \bibinfo{year}{2019}\natexlab{}.
\newblock \showarticletitle{Causal confusion in imitation learning}.
\newblock \bibinfo{journal}{\emph{Advances in Neural Information Processing Systems}}  \bibinfo{volume}{32} (\bibinfo{year}{2019}).
\newblock


\bibitem[Deng et~al\mbox{.}(2020)]%
        {DENG2020101656}
\bibfield{author}{\bibinfo{person}{Changyu Deng}, \bibinfo{person}{Xunbi Ji}, \bibinfo{person}{Colton Rainey}, \bibinfo{person}{Jianyu Zhang}, {and} \bibinfo{person}{Wei Lu}.} \bibinfo{year}{2020}\natexlab{}.
\newblock \showarticletitle{Integrating Machine Learning with Human Knowledge}.
\newblock \bibinfo{journal}{\emph{iScience}} \bibinfo{volume}{23}, \bibinfo{number}{11} (\bibinfo{year}{2020}), \bibinfo{pages}{101656}.
\newblock


\bibitem[Fung et~al\mbox{.}(2002)]%
        {fung2002knowledge}
\bibfield{author}{\bibinfo{person}{Glenn Fung}, \bibinfo{person}{Olvi Mangasarian}, {and} \bibinfo{person}{Jude Shavlik}.} \bibinfo{year}{2002}\natexlab{}.
\newblock \showarticletitle{Knowledge-based support vector machine classifiers}. In \bibinfo{booktitle}{\emph{NeurIPS}}.
\newblock


\bibitem[Garc{\i}a and Fern{\'a}ndez(2015)]%
        {garcia2015comprehensive}
\bibfield{author}{\bibinfo{person}{Javier Garc{\i}a} {and} \bibinfo{person}{Fernando Fern{\'a}ndez}.} \bibinfo{year}{2015}\natexlab{}.
\newblock \showarticletitle{A comprehensive survey on safe reinforcement learning}.
\newblock \bibinfo{journal}{\emph{Journal of Machine Learning Research}} \bibinfo{volume}{16}, \bibinfo{number}{1} (\bibinfo{year}{2015}), \bibinfo{pages}{1437--1480}.
\newblock


\bibitem[Goldberger and Ben-Reuven(2017)]%
        {goldberger2016training}
\bibfield{author}{\bibinfo{person}{Jacob Goldberger} {and} \bibinfo{person}{Ehud Ben-Reuven}.} \bibinfo{year}{2017}\natexlab{}.
\newblock \showarticletitle{Training deep neural-networks using a noise adaptation layer}. In \bibinfo{booktitle}{\emph{ICLR}}.
\newblock


\bibitem[Hadary et~al\mbox{.}(2020)]%
        {hadary2020protean}
\bibfield{author}{\bibinfo{person}{Ori Hadary}, \bibinfo{person}{Luke Marshall}, \bibinfo{person}{Ishai Menache}, \bibinfo{person}{Abhisek Pan}, \bibinfo{person}{Esaias~E Greeff}, \bibinfo{person}{David Dion}, \bibinfo{person}{Star Dorminey}, \bibinfo{person}{Shailesh Joshi}, \bibinfo{person}{Yang Chen}, \bibinfo{person}{Mark Russinovich}, {et~al\mbox{.}}} \bibinfo{year}{2020}\natexlab{}.
\newblock \showarticletitle{Protean:{VM} Allocation Service at Scale}. In \bibinfo{booktitle}{\emph{14th {USENIX} Symposium on Operating Systems Design and Implementation ({OSDI} 20)}}. \bibinfo{pages}{845--861}.
\newblock


\bibitem[Harel and Radinsky(2018)]%
        {harel2018accelerating}
\bibfield{author}{\bibinfo{person}{Shahar Harel} {and} \bibinfo{person}{Kira Radinsky}.} \bibinfo{year}{2018}\natexlab{}.
\newblock \showarticletitle{Accelerating prototype-based drug discovery using conditional diversity networks}. In \bibinfo{booktitle}{\emph{Proceedings of the 24th ACM SIGKDD International Conference on Knowledge Discovery \& Data Mining}}. \bibinfo{pages}{331--339}.
\newblock


\bibitem[Ho and Ermon(2016)]%
        {ho2016generative}
\bibfield{author}{\bibinfo{person}{Jonathan Ho} {and} \bibinfo{person}{Stefano Ermon}.} \bibinfo{year}{2016}\natexlab{}.
\newblock \showarticletitle{Generative adversarial imitation learning}. In \bibinfo{booktitle}{\emph{NeurIPS}}, Vol.~\bibinfo{volume}{29}.
\newblock


\bibitem[Householder et~al\mbox{.}(2014)]%
        {householder2014cloud}
\bibfield{author}{\bibinfo{person}{Rachel Householder}, \bibinfo{person}{Scott Arnold}, {and} \bibinfo{person}{Robert Green}.} \bibinfo{year}{2014}\natexlab{}.
\newblock \showarticletitle{On cloud-based oversubscription}.
\newblock \bibinfo{journal}{\emph{arXiv preprint arXiv:1402.4758}} (\bibinfo{year}{2014}).
\newblock


\bibitem[Hussein et~al\mbox{.}(2017)]%
        {hussein2017imitation}
\bibfield{author}{\bibinfo{person}{Ahmed Hussein}, \bibinfo{person}{Mohamed~Medhat Gaber}, \bibinfo{person}{Eyad Elyan}, {and} \bibinfo{person}{Chrisina Jayne}.} \bibinfo{year}{2017}\natexlab{}.
\newblock \showarticletitle{Imitation learning: A survey of learning methods}.
\newblock \bibinfo{journal}{\emph{ACM Computing Surveys (CSUR)}} \bibinfo{volume}{50}, \bibinfo{number}{2} (\bibinfo{year}{2017}), \bibinfo{pages}{1--35}.
\newblock


\bibitem[Jiang et~al\mbox{.}(2018)]%
        {jiang2018mentornet}
\bibfield{author}{\bibinfo{person}{Lu Jiang}, \bibinfo{person}{Zhengyuan Zhou}, \bibinfo{person}{Thomas Leung}, \bibinfo{person}{Li-Jia Li}, {and} \bibinfo{person}{Li Fei-Fei}.} \bibinfo{year}{2018}\natexlab{}.
\newblock \showarticletitle{Mentornet: Learning data-driven curriculum for very deep neural networks on corrupted labels}. In \bibinfo{booktitle}{\emph{ICML}}.
\newblock


\bibitem[Kim et~al\mbox{.}(2016)]%
        {kim2016examples}
\bibfield{author}{\bibinfo{person}{Been Kim}, \bibinfo{person}{Rajiv Khanna}, {and} \bibinfo{person}{Oluwasanmi~O Koyejo}.} \bibinfo{year}{2016}\natexlab{}.
\newblock \showarticletitle{Examples are not enough, learn to criticize! criticism for interpretability}. In \bibinfo{booktitle}{\emph{Advances in neural information processing systems}}. \bibinfo{pages}{2280--2288}.
\newblock


\bibitem[Kunapuli et~al\mbox{.}(2013)]%
        {kunapuli2013guiding}
\bibfield{author}{\bibinfo{person}{Gautam Kunapuli}, \bibinfo{person}{Phillip Odom}, \bibinfo{person}{Jude~W Shavlik}, {and} \bibinfo{person}{Sriraam Natarajan}.} \bibinfo{year}{2013}\natexlab{}.
\newblock \showarticletitle{Guiding autonomous agents to better behaviors through human advice}. In \bibinfo{booktitle}{\emph{ICDM}}.
\newblock


\bibitem[Lee(2013)]%
        {lee2013pseudo}
\bibfield{author}{\bibinfo{person}{Dong-Hyun Lee}.} \bibinfo{year}{2013}\natexlab{}.
\newblock \showarticletitle{Pseudo-label: The simple and efficient semi-supervised learning method for deep neural networks}. In \bibinfo{booktitle}{\emph{Workshop on Challenges in Representation Learning, ICML}}.
\newblock


\bibitem[Li et~al\mbox{.}(2010)]%
        {li2010survey}
\bibfield{author}{\bibinfo{person}{Yunfa Li}, \bibinfo{person}{Wanqing Li}, {and} \bibinfo{person}{Congfeng Jiang}.} \bibinfo{year}{2010}\natexlab{}.
\newblock \showarticletitle{A survey of virtual machine system: Current technology and future trends}. In \bibinfo{booktitle}{\emph{2010 Third International Symposium on Electronic Commerce and Security}}. IEEE, \bibinfo{pages}{332--336}.
\newblock


\bibitem[Li(2019)]%
        {li2019adaptive}
\bibfield{author}{\bibinfo{person}{Zhihua Li}.} \bibinfo{year}{2019}\natexlab{}.
\newblock \showarticletitle{An adaptive overload threshold selection process using Markov decision processes of virtual machine in cloud data center}.
\newblock \bibinfo{journal}{\emph{Cluster Computing}} \bibinfo{volume}{22}, \bibinfo{number}{2} (\bibinfo{year}{2019}), \bibinfo{pages}{3821--3833}.
\newblock


\bibitem[Lillicrap et~al\mbox{.}(2016)]%
        {lillicrap2016continuous}
\bibfield{author}{\bibinfo{person}{Timothy~P Lillicrap}, \bibinfo{person}{Jonathan~J Hunt}, \bibinfo{person}{Alexander Pritzel}, \bibinfo{person}{Nicolas Heess}, \bibinfo{person}{Tom Erez}, \bibinfo{person}{Yuval Tassa}, \bibinfo{person}{David Silver}, {and} \bibinfo{person}{Daan Wierstra}.} \bibinfo{year}{2016}\natexlab{}.
\newblock \showarticletitle{Continuous control with deep reinforcement learning.}. In \bibinfo{booktitle}{\emph{ICLR}}.
\newblock


\bibitem[Mahapatra and Venkatrao(1999)]%
        {mahapatra1999processor}
\bibfield{author}{\bibinfo{person}{Nihar~R Mahapatra} {and} \bibinfo{person}{Balakrishna Venkatrao}.} \bibinfo{year}{1999}\natexlab{}.
\newblock \showarticletitle{The processor-memory bottleneck: problems and solutions}.
\newblock \bibinfo{journal}{\emph{Crossroads}} \bibinfo{volume}{5}, \bibinfo{number}{3es} (\bibinfo{year}{1999}), \bibinfo{pages}{2}.
\newblock


\bibitem[Mazyavkina et~al\mbox{.}(2021)]%
        {mazyavkina2021reinforcement}
\bibfield{author}{\bibinfo{person}{Nina Mazyavkina}, \bibinfo{person}{Sergey Sviridov}, \bibinfo{person}{Sergei Ivanov}, {and} \bibinfo{person}{Evgeny Burnaev}.} \bibinfo{year}{2021}\natexlab{}.
\newblock \showarticletitle{Reinforcement learning for combinatorial optimization: A survey}.
\newblock \bibinfo{journal}{\emph{Computers \& Operations Research}}  \bibinfo{volume}{134} (\bibinfo{year}{2021}), \bibinfo{pages}{105400}.
\newblock


\bibitem[Ming et~al\mbox{.}(2019)]%
        {ming2019interpretable}
\bibfield{author}{\bibinfo{person}{Yao Ming}, \bibinfo{person}{Panpan Xu}, \bibinfo{person}{Huamin Qu}, {and} \bibinfo{person}{Liu Ren}.} \bibinfo{year}{2019}\natexlab{}.
\newblock \showarticletitle{Interpretable and steerable sequence learning via prototypes}. In \bibinfo{booktitle}{\emph{Proceedings of the 25th ACM SIGKDD International Conference on Knowledge Discovery \& Data Mining}}. \bibinfo{pages}{903--913}.
\newblock


\bibitem[Minton et~al\mbox{.}(1989)]%
        {minton1989explanation}
\bibfield{author}{\bibinfo{person}{Steven Minton}, \bibinfo{person}{Jaime~G Carbonell}, \bibinfo{person}{Craig~A Knoblock}, \bibinfo{person}{Daniel~R Kuokka}, \bibinfo{person}{Oren Etzioni}, {and} \bibinfo{person}{Yolanda Gil}.} \bibinfo{year}{1989}\natexlab{}.
\newblock \showarticletitle{Explanation-based learning: A problem solving perspective}.
\newblock \bibinfo{journal}{\emph{Artificial Intelligence}} \bibinfo{volume}{40}, \bibinfo{number}{1-3} (\bibinfo{year}{1989}), \bibinfo{pages}{63--118}.
\newblock


\bibitem[Miyato et~al\mbox{.}(2018)]%
        {miyato2018virtual}
\bibfield{author}{\bibinfo{person}{Takeru Miyato}, \bibinfo{person}{Shin-ichi Maeda}, \bibinfo{person}{Shin Ishii}, {and} \bibinfo{person}{Masanori Koyama}.} \bibinfo{year}{2018}\natexlab{}.
\newblock \showarticletitle{Virtual adversarial training: a regularization method for supervised and semi-supervised learning}.
\newblock \bibinfo{journal}{\emph{TPAMI}} (\bibinfo{year}{2018}).
\newblock


\bibitem[Molnar(2020)]%
        {molnar2020interpretable}
\bibfield{author}{\bibinfo{person}{Christoph Molnar}.} \bibinfo{year}{2020}\natexlab{}.
\newblock \bibinfo{booktitle}{\emph{Interpretable machine learning}}.
\newblock \bibinfo{publisher}{Lulu. com}.
\newblock


\bibitem[Nazifi et~al\mbox{.}(2021)]%
        {nazifi2021proactive}
\bibfield{author}{\bibinfo{person}{Amin Nazifi}, \bibinfo{person}{Katja Gelbrich}, \bibinfo{person}{Yany Gr{\'e}goire}, \bibinfo{person}{Sebastian Koch}, \bibinfo{person}{Dahlia El-Manstrly}, {and} \bibinfo{person}{Jochen Wirtz}.} \bibinfo{year}{2021}\natexlab{}.
\newblock \showarticletitle{Proactive handling of flight overbooking: how to reduce negative eWOM and the costs of bumping customers}.
\newblock \bibinfo{journal}{\emph{Journal of Service Research}} \bibinfo{volume}{24}, \bibinfo{number}{2} (\bibinfo{year}{2021}), \bibinfo{pages}{206--225}.
\newblock


\bibitem[Neider et~al\mbox{.}(2021)]%
        {neider2021advice}
\bibfield{author}{\bibinfo{person}{Daniel Neider}, \bibinfo{person}{Jean-Raphael Gaglione}, \bibinfo{person}{Ivan Gavran}, \bibinfo{person}{Ufuk Topcu}, \bibinfo{person}{Bo Wu}, {and} \bibinfo{person}{Zhe Xu}.} \bibinfo{year}{2021}\natexlab{}.
\newblock \showarticletitle{Advice-guided reinforcement learning in a non-markovian environment}. In \bibinfo{booktitle}{\emph{AAAI}}.
\newblock


\bibitem[Odom and Natarajan(2015)]%
        {odom2015active}
\bibfield{author}{\bibinfo{person}{Phillip Odom} {and} \bibinfo{person}{Sriraam Natarajan}.} \bibinfo{year}{2015}\natexlab{}.
\newblock \showarticletitle{Active advice seeking for inverse reinforcement learning}. In \bibinfo{booktitle}{\emph{AAAI}}.
\newblock


\bibitem[Paternain et~al\mbox{.}(2019)]%
        {paternain2019constrained}
\bibfield{author}{\bibinfo{person}{Santiago Paternain}, \bibinfo{person}{Luiz Chamon}, \bibinfo{person}{Miguel Calvo-Fullana}, {and} \bibinfo{person}{Alejandro Ribeiro}.} \bibinfo{year}{2019}\natexlab{}.
\newblock \showarticletitle{Constrained reinforcement learning has zero duality gap}.
\newblock \bibinfo{journal}{\emph{Advances in Neural Information Processing Systems}}  \bibinfo{volume}{32} (\bibinfo{year}{2019}).
\newblock


\bibitem[Patrini et~al\mbox{.}(2017)]%
        {patrini2017making}
\bibfield{author}{\bibinfo{person}{Giorgio Patrini}, \bibinfo{person}{Alessandro Rozza}, \bibinfo{person}{Aditya~Krishna Menon}, \bibinfo{person}{Richard Nock}, {and} \bibinfo{person}{Lizhen Qu}.} \bibinfo{year}{2017}\natexlab{}.
\newblock \showarticletitle{Making deep neural networks robust to label noise: A loss correction approach}. In \bibinfo{booktitle}{\emph{CVPR}}.
\newblock


\bibitem[Pomerleau(1991)]%
        {pomerleau1991efficient}
\bibfield{author}{\bibinfo{person}{Dean~A Pomerleau}.} \bibinfo{year}{1991}\natexlab{}.
\newblock \showarticletitle{Efficient training of artificial neural networks for autonomous navigation}.
\newblock \bibinfo{journal}{\emph{Neural computation}} (\bibinfo{year}{1991}).
\newblock


\bibitem[Qiao et~al\mbox{.}(2021)]%
        {qiao2021intelligent}
\bibfield{author}{\bibinfo{person}{Bo Qiao}, \bibinfo{person}{Fangkai Yang}, \bibinfo{person}{Chuan Luo}, \bibinfo{person}{Yanan Wang}, \bibinfo{person}{Johnny Li}, \bibinfo{person}{Qingwei Lin}, \bibinfo{person}{Hongyu Zhang}, \bibinfo{person}{Mohit Datta}, \bibinfo{person}{Andrew Zhou}, \bibinfo{person}{Thomas Moscibroda}, {et~al\mbox{.}}} \bibinfo{year}{2021}\natexlab{}.
\newblock \showarticletitle{Intelligent container reallocation at Microsoft 365}. In \bibinfo{booktitle}{\emph{Proceedings of the 29th ACM Joint Meeting on European Software Engineering Conference and Symposium on the Foundations of Software Engineering}}. \bibinfo{pages}{1438--1443}.
\newblock


\bibitem[Rosch(1973)]%
        {rosch1973natural}
\bibfield{author}{\bibinfo{person}{Eleanor~H Rosch}.} \bibinfo{year}{1973}\natexlab{}.
\newblock \showarticletitle{Natural categories}.
\newblock \bibinfo{journal}{\emph{Cognitive psychology}} \bibinfo{volume}{4}, \bibinfo{number}{3} (\bibinfo{year}{1973}), \bibinfo{pages}{328--350}.
\newblock


\bibitem[Suryani et~al\mbox{.}(2010)]%
        {suryani2010air}
\bibfield{author}{\bibinfo{person}{Erma Suryani}, \bibinfo{person}{Shuo-Yan Chou}, {and} \bibinfo{person}{Chih-Hsien Chen}.} \bibinfo{year}{2010}\natexlab{}.
\newblock \showarticletitle{Air passenger demand forecasting and passenger terminal capacity expansion: A system dynamics framework}.
\newblock \bibinfo{journal}{\emph{Expert Systems with Applications}} \bibinfo{volume}{37}, \bibinfo{number}{3} (\bibinfo{year}{2010}), \bibinfo{pages}{2324--2339}.
\newblock


\bibitem[Suzuki(2006)]%
        {suzuki2006net}
\bibfield{author}{\bibinfo{person}{Yoshinori Suzuki}.} \bibinfo{year}{2006}\natexlab{}.
\newblock \showarticletitle{The net benefit of airline overbooking}.
\newblock \bibinfo{journal}{\emph{Transportation Research Part E: Logistics and Transportation Review}} \bibinfo{volume}{42}, \bibinfo{number}{1} (\bibinfo{year}{2006}), \bibinfo{pages}{1--19}.
\newblock


\bibitem[Wang and Tianfield(2018)]%
        {wang2018energy}
\bibfield{author}{\bibinfo{person}{Hui Wang} {and} \bibinfo{person}{Huaglory Tianfield}.} \bibinfo{year}{2018}\natexlab{}.
\newblock \showarticletitle{Energy-aware dynamic virtual machine consolidation for cloud datacenters}.
\newblock \bibinfo{journal}{\emph{IEEE Access}}  \bibinfo{volume}{6} (\bibinfo{year}{2018}), \bibinfo{pages}{15259--15273}.
\newblock


\bibitem[Williams et~al\mbox{.}(2011)]%
        {williams2011overdriver}
\bibfield{author}{\bibinfo{person}{Dan Williams}, \bibinfo{person}{Hani Jamjoom}, \bibinfo{person}{Yew-Huey Liu}, {and} \bibinfo{person}{Hakim Weatherspoon}.} \bibinfo{year}{2011}\natexlab{}.
\newblock \showarticletitle{Overdriver: Handling memory overload in an oversubscribed cloud}.
\newblock \bibinfo{journal}{\emph{ACM SIGPLAN Notices}} \bibinfo{volume}{46}, \bibinfo{number}{7} (\bibinfo{year}{2011}), \bibinfo{pages}{205--216}.
\newblock


\bibitem[Wittman(2014)]%
        {wittman2014low}
\bibfield{author}{\bibinfo{person}{Michael~D Wittman}.} \bibinfo{year}{2014}\natexlab{}.
\newblock \showarticletitle{Are low-cost carrier passengers less likely to complain about service quality?}
\newblock \bibinfo{journal}{\emph{Journal of Air Transport Management}}  \bibinfo{volume}{35} (\bibinfo{year}{2014}), \bibinfo{pages}{64--71}.
\newblock


\end{thebibliography}

\end{document}